\documentclass[twoside,11pt]{article}

\usepackage{blindtext}
\usepackage{ulem}
\usepackage[preprint]{jmlr2e}

\usepackage{lastpage}

\usepackage{xcolor}      
\usepackage{minted}      
\usepackage{subcaption}

\definecolor{codebg}{RGB}{248,248,248}  

\setminted{
  bgcolor=codebg,
  frame=none,           
  framerule=0pt,        
  framesep=6pt,         
  fontsize=\small,
  breaklines=true
}

\usepackage{amsmath}
\usepackage{bm}
\usepackage{listings}
\usepackage{algorithm}
\usepackage{algpseudocode}

\jmlrheading{23}{2022}{1-\pageref{LastPage}}{1/21; Revised 5/22}{9/22}{21-0000}{Martin Tveten and Johannes Voll Kolstø}

\ShortHeadings{skchange}{Tveten, Kolstø, Moen}
\firstpageno{1}

\begin{document}

\title{skchange: Fast and Flexible Algorithms for Changepoint Detection}

\author{\\
    \name Martin Tveten$^{1}$ \hfill \email tveten@nr.no\\
    \name Johannes Voll Kolstø$^{1}$ \hfill \email jvkolsto@nr.no\\
    \name Per August Jarval Moen$^{2}$ \hfill \email pamoen@math.uio.no\\
    \addr $^{1}$ Department of Statistics and Machine Learning, Norwegian Computing Center\\
    \addr $^{2}$ Department of Mathematics, University of Oslo
}

\editor{My editor}

\maketitle

\begin{abstract}
Skchange is an open-source Python library for detecting structural changes in time series.
It implements modern change detection algorithms within a unified and extensible framework.
The algorithms are modular and composable, and they include changepoint search methods based on both cost minimisation and statistical tests.
Key features include
the detection of anomalous segments in addition to changepoints;
theoretically well-founded fast and approximate search methods;
theoretically well-founded algorithms for high-dimensional data, covering settings where either few or many features change simultaneously;
utilities for automatic and data-driven penalty calibration, which balances false alarms against missed detections;
and a large collection of built-in costs and statistical tests.
The design follows established scikit-learn conventions to streamline both user and contributor experience, and Numba is used extensively to achieve high computational performance.
Source code and documentation are available at \url{https://github.com/NorskRegnesentral/skchange}.
\end{abstract}


\section{Introduction}
\label{sec:intro}
\thispagestyle{empty}

\begin{figure}[t]
    \centering
    \captionsetup[subfigure]{skip=0pt, belowskip=0pt}
    \begin{subfigure}[t]{0.49\linewidth}
        \centering
        \includegraphics[width=\linewidth]{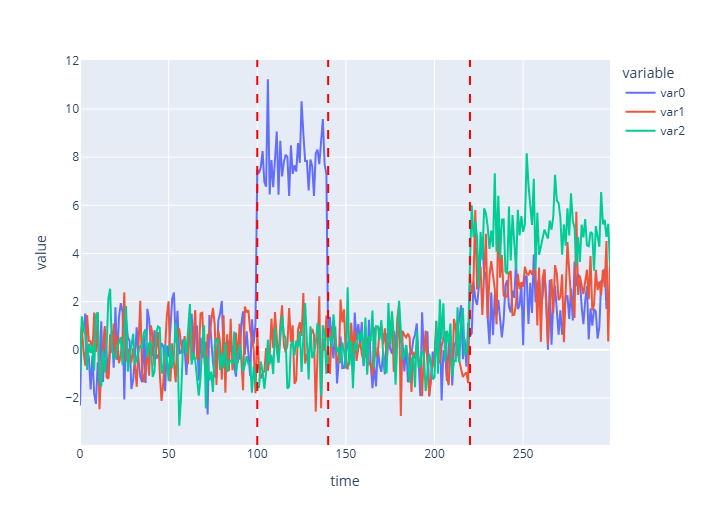}
        \caption{}
        \label{fig:data_changepoints}
    \end{subfigure}
    \hfill
    \begin{subfigure}[t]{0.49\linewidth}
        \centering
        \includegraphics[width=\linewidth]{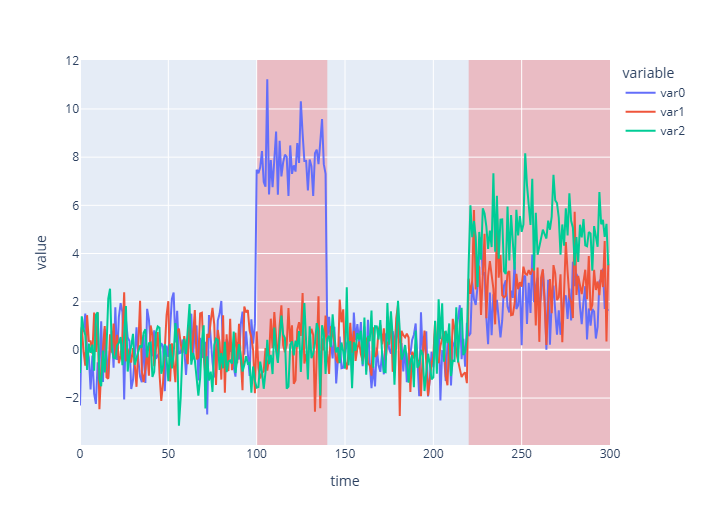}
        \caption{}
        \label{fig:data_anomalies}
    \end{subfigure}
    \caption{Three-dimensional toy data with (a) three detected changepoints and (b) two detected segment anomalies as an alternative representation.}

    \label{fig:data_example}
\end{figure}

Detecting structural changes in time series is a fundamental task in many machine learning applications, including personalised medicine \citep{li2026change}, condition monitoring of industrial equipment \citep{tveten2022scalable}, environmental monitoring \citep{gong2025changepoint}, and fraud detection \citep{rousseeuw2019robust}.
The key problem is to detect the presence of \emph{changepoints} and estimate their locations, where a changepoint marks an abrupt change in the statistical properties of a sequence (Figure~\ref{fig:data_changepoints}).

Recent advances in the statistics community have produced algorithms that are computationally efficient, theoretically well-founded, and flexible with respect to data-generating mechanisms.
These methods are interpretable and can be tailored to specific domains.
However, despite their practical relevance, most of these advances are not available in a user-friendly form in Python, limiting their adoption by machine learning practitioners.

In this paper, we introduce skchange, an open-source Python library for changepoint detection that aims to be both computationally efficient and easy to use and extend.
We have developed an API for changepoint detection that follows scikit-learn conventions \citep{pedregosa2011scikit} closely.
Algorithms are implemented as modular compositions, and high computational performance is achieved through Numba \citep{lam2015numba} for just-in-time compilation.

The closest previous work is the ruptures library \citep{truong2020selective}, which compiles several classical changepoint detection algorithms and provides a wide range of cost functions to be optimised.
Compared to ruptures, the scope of skchange is broader.
Firstly, it does not restrict algorithms to cost-minimisation frameworks, enabling implementation of recent advances based on statistical tests.
Among other advantages, such test-based methods are more suitable for high-dimensional data and can be more efficient to calibrate for a desired false alarm rate.
Secondly, skchange supports segment anomaly detection, which can be viewed as a changepoint detection problem where the aim is to find segments that deviate from some baseline behaviour (Figure~\ref{fig:data_anomalies}).
In addition, skchange features improved performance (see Section~\ref{sec:features}).

\section{Design}
\label{sec:design}

\paragraph{Scikit-learn interface.}
Skchange follows scikit-learn estimator conventions, including \texttt{fit} and \texttt{predict} semantics, and the separation of hyperparameters from learned attributes.
The key difference from other scikit-learn estimators is that \texttt{predict} returns detected changepoints as integer sample indices.
Figure~\ref{fig:code_example} shows a code example of detecting changepoints using skchange.
This common interface supports interchangeable detector use and enables detector-agnostic tools for preprocessing, hyperparameter tuning and evaluation.

\paragraph{Composable detectors.}
Detectors in skchange are composed of three core components: a search algorithm, an interval scorer, and a penalty.
In skchange, we introduce \emph{interval scorer} as an abstraction for cost functions and test statistics applied to interval subsets of a dataset.
Specifically, an interval scorer maps a dataset and a set of \emph{interval specifications} to a value quantifying evidence for change or model deviation within the interval. By an interval specification, we refer to the indices encoding the interval and, for two-sample test statistics, the sample splitting information.
The search algorithm determines which intervals to assess and how to combine results into a final set of detected changepoints.
In Figure~\ref{fig:code_example}, \texttt{CUSUM} is the interval scorer and \texttt{SeededBinarySegmentation} is the search algorithm.
The penalty controls model complexity by regulating the number of detected events.
This modular design allows new detectors to be constructed by combining existing components or introducing new ones without altering the surrounding workflow.

\paragraph{Vectorised interval scorers with precomputation.}
The computational bottleneck of most change detection algorithms is to compute interval scores over a large number of interval specifications, often with overlapping regions.
Our design of interval scorers provides two mechanisms for speeding this up.
Firstly, they can precompute quantities that are reused across many interval specifications, such as cumulative sums and other sufficient statistics.
Secondly, they are vectorised over interval specifications, enabling batch evaluation of large candidate sets with Numba-accelerated implementations.

\begin{figure}[t]
\centering
\begin{minted}[fontsize=\footnotesize]{python}
from skchange.detectors import SeededBinarySegmentation
from skchange.interval_scorers import CUSUM
score = CUSUM()
detector = SeededBinarySegmentation(score, penalty=5)
detector.fit(X)
changepoints = detector.predict(X)
\end{minted}
\caption{
Code example of detecting changepoints in an array \texttt{X} using Seeded Binary Segmentation with a CUSUM score. Skchange also offer tools for calibrating the \texttt{penalty}.
}
\label{fig:code_example}
\end{figure}

\section{Features}
\label{sec:features}
This section summarises the features that distinguish skchange from other packages.

\paragraph{State-of-the-art changepoint search methods.}
Supported methods include the moving window or MOSUM algorithm \citep{eichinger2018mosum, meier2021mosum}, generalised to arbitrary test statistics and allowing multiple bandwidths and different selection methods;
Seeded Binary Segmentation \citep{kovacs2023seeded}, with optional narrowest-over-threshold selection \citep{baranowski2019narrowest}; PELT \citep{killick2012optimal}; FPOP \citep{maidstone2017optimal}; and CROPS \citep{haynes2017crops}.
Of these, only PELT and a basic version of moving window are also available in ruptures.
Seeded Binary Segmentation and the moving window algorithm are fast test-based algorithms with strong theoretical guarantees.

\paragraph{Segment anomaly detection.}
Skchange implements two change detection algorithms that also support segment anomaly detection:
CAPA and its multivariate extensions \citep{fisch2022capa, fisch2022mvcapa}, and Circular Binary Segmentation \citep{olshen2004circular}.
For multivariate data, CAPA additionally identifies which variables are affected by an anomaly.

\paragraph{Wide range of efficient scorers.}
The package provides a broad set of interval scorers covering common modelling assumptions and distributional features.
Nearly all cost functions in ruptures are reimplemented for higher performance using the vectorised scorer design combined with Numba.
These include parametric likelihood-based costs, trend and regression costs, and nonparametric measures such as rank-based costs.
In addition, skchange offers efficient implementations of scores based on statistical tests for changes in the mean (CUSUM), continuous linear trends, multivariate $t$-distributions, and more.

\paragraph{High-dimensional data.}
In high-dimensional settings, changes may be \emph{sparse} (affecting few variables) or \emph{dense} (affecting many). Classical statistical tests typically have power for only one regime.
Skchange includes scores based on recent statistical tests that are sensitive to both regimes while remaining computationally efficient, such as ESAC \citep{moen2024esac} and SUBSET \citep{tickle2021computationally}.

\paragraph{Automatic penalty calibration.}
As the penalty governs the trade-off between false alarms and missed detections, calibrating the penalty parameter of change detectors is a critical task before the detector can be used reliably in real settings.
Skchange features data-driven tools for automatically calibrating the penalty to control the family-wise error rate.
Examples include methods based on permutations or the block bootstrap.

\begin{figure}[t]
    \centering
    \includegraphics[width=1.0\textwidth]{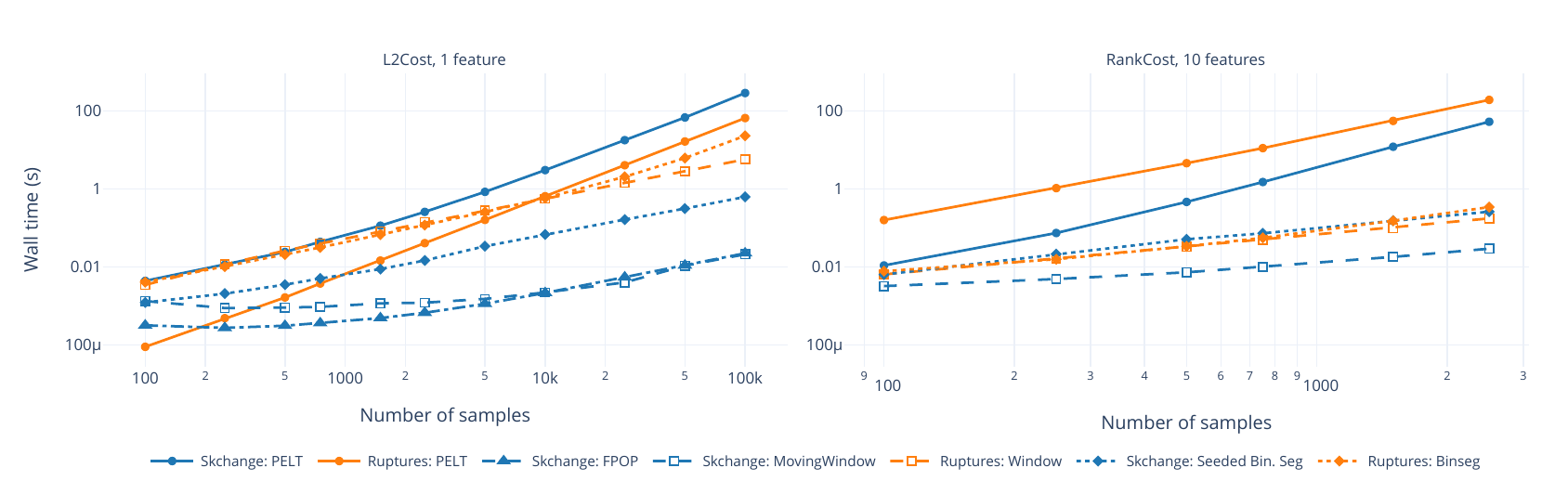}
    \caption{Run-time evaluation of comparable algorithms in skchange and ruptures on change-free Gaussian data for the L2 cost (left) and the multivariate rank cost (right). Machine specifications: CPU Intel Xeon Silver 4110 @ 2.10 GHz, 10 cores, 49 GB of RAM.
    }
    \label{fig:runtime_evaluation}
\end{figure}

\paragraph{Run-time efficiency.}
Skchange is notably faster than ruptures in most cases.
Figure~\ref{fig:runtime_evaluation} showcases run-time evaluations of comparable algorithms in skchange and ruptures for the canonical L2 cost for detecting abrupt changes in the mean in univariate data, and the non-parametric rank cost.
The evaluation was performed on change-free Gaussian data for simplicity.
For the L2 cost example, please note that FPOP and PELT are comparable as they solve the same optimisation problem, and that we use ruptures' C-implementation of PELT in their \texttt{KernelCPD} detector.
A more extensive set of run-time evaluations is available at \url{https://github.com/NorskRegnesentral/change-point-benchmark}.

\section{Contribution and Development Practices}
\label{sec:contributions}
Skchange is released under the BSD 3-Clause license and is publicly available at \url{https://github.com/NorskRegnesentral/skchange}.
Comprehensive documentation is available at \href{https://skchange.readthedocs.io}{\url{https://skchange.readthedocs.io}}, including user and developer guides, and a complete API reference.
Reliability is ensured through near-complete test coverage and continuous integration across Python versions and platforms, including Linux and Windows.
The library adheres to established software development best practices and has been deployed in production systems by industrial partners.
Further contributions from students, researchers and the open-source community are highly welcome.


\acks{This project was supported by Norwegian Research Council grants 332645 (Integreat), 337085 (EarOnEdge) and 356322 (SODA).
Various AI tools have been used for proofreading and language improvements of the manuscript, including Microsoft Copilot and Claude Sonnet 4.5 and 4.6.
The same AI tools have also been used to iterate on the software code since 2025.
Thanks to Franz J. Király and the sktime community for valuable discussions during the early development of this work.
An earlier version of skchange is available as part of the sktime library \citep{loning2019sktime}.
}

\vskip 0.2in
\bibliography{main}

\end{document}